\documentclass[reprint,secnumarabic, amssymb, amsmath,aip,cha,groupedaddress,frontmatterverbose]{revtex4-1}  

\usepackage{graphicx}
\usepackage{amsmath,amssymb,amsfonts,mathtools}
\usepackage{ulem}
\usepackage{color}
\usepackage{multirow}
\usepackage{float}
\usepackage{soul}
\begin{document}

\title{Investigating Photoinduced Proton Coupled Electron Transfer Reaction using Quasi Diabatic Dynamics Propagation}
\author{Arkajit Mandal}%
\author{Farnaz A. Shakib}
\author{Pengfei Huo}%
\email{pengfei.huo@rochester.edu}
\affiliation{Department of Chemistry, University of Rochester, 120 Trustee Road, Rochester, New York 14627, United States}%

\begin{abstract}
We investigate photoinduced proton-coupled electron transfer (PI-PCET) reaction through a recently developed quasi-diabatic (QD) quantum dynamics propagation scheme. This scheme enables interfacing accurate diabatic-based quantum dynamics approaches with adiabatic electronic structure calculations for on-the-fly simulations. Here, we use the QD scheme to directly propagate PI-PCET quantum dynamics with the diabatic Partial Linearized Density Matrix (PLDM) path-integral approach with the instantaneous adiabatic electron-proton vibronic states. Our numerical results demonstrate the importance of treating proton quantum mechanically in order to obtain accurate PI-PCET dynamics, as well as the role of solvent fluctuation and vibrational relaxation on proton tunneling in various reaction regimes that exhibit different kinetic isotope effects. This work opens the possibility to study the challenging PI-PCET reactions through accurate diabatic quantum dynamics approaches combined with efficient adiabatic electronic structure calculations. 
\end{abstract}

\maketitle
 
\section{Introduction}
Photoinduced proton-coupled electron transfer (PI-PCET) reactions\cite{Gagliardi:2010,Goyal:2017,Lennox:2017} involve the coupled transfer of both electron and proton upon photoexcitation. Thus, PI-PCET is fundamentally different from the extensively studied photoinduced proton transfer (PT) or electron transfer (ET) reactions. Several recent experimental and theoretical studies have revealed PI-PCET reactions in a wide range of systems,\cite{Lennox:2017} such as simple hydrogen-bonded organic complexes,\cite{Westlake:2011,Thompson:2016, Eisenhart:2014,Goyal:2017} organometallic complexes,\cite{Hodgkiss:2006,Rosenthal:2006,Young:2009} enol-keto tautomerization,\cite{Luber:2013,Kim:2009} as well as photocatalytic water oxidation on small nanoparticles.\cite{Muuronen:2017}

Initiated through photoexcitation process,  PI-PCET reactions play a critical role in solar energy conversion processes.\cite{Gust:2009,Lennox:2017} At the same time, they are promising for providing new and unique reactivities which are not directly accessible in regular thermally activated PCET reactions.\cite{Goyal:2017,Lennox:2017} Thus, understanding the fundamental mechanistic principles of PI-PCET will allow tuning and controlling this reaction and provide design principles for more efficient solar energy conversion devices. 

Accurately and efficiently simulating PI-PCET reactions, however, remains a challenging theoretical task as it requires an explicit quantum mechanical description of both electronic nonadiabatic transitions and nuclear quantum effects for proton (such as tunneling and zero-point energy). Further, the charge distribution in the excited state can significantly deviate from the ground state,\cite{Goyal:2017} leading to a highly non-equilibrium configuration of the solvent upon initial photoexcitaiton. Relaxation of the solvent initial configuration can then drastically affect the reaction course of the PI-PCET.\cite{Hazra:2010,Hazra:2011,Goyal:2015} Thus, compared to the well-explored thermal PCET reactions,\cite{Fang:1997,Cukier:1998,Soudackov:2000,Hammes-Schiffer:2001,Soudackov:2005,Huynh:2007,Hammes-Schiffer:2009,Hammes-Schiffer:2010,Weinberg:2012,Migliore:2014,shakib:2015,shakib:2016,Ananth:2012, Kretchmer:2013,Kretchmer:2016, Ananth:2017} the proper description of the non-equilibrium PI-PCET process is beyond the equilibrium rate constant expressions\cite{Hammes-Schiffer2012} and requires either a detailed time-dependent dynamics\cite{Hazra:2010,Hazra:2011,Goyal:2015,Goyal:2016} or non-equilibrium Fermi's golden rule.\cite{Shi:2017,Geva:2016}

Recent theoretical studies based on fewest-switches surface-hopping (FSSH)\cite{Tully:1990,Hammes-schiffer:1994} simulations with electron-proton vibronic basis have made significant contributions to elucidating PI-PCET dynamics.\cite{Hazra:2010,Hazra:2011,Goyal:2015,Goyal:2016} As a mixed quantum-classical (MQC) method, however, FSSH treats quantum and classical degrees of freedom (DOF) on different footings. This can generate artificial coherence that leads to incorrect ET dynamics\cite{Landry:2011} or the breakdown of the detailed balance.\cite{Schmidt:2008} Several modified FSSH methods,\cite{Landry:2011,Granucci:2010,Jaeger:2012,Akimov:2014,Wang:2015a,Wang:2015b} especially decoherence corrected FSSH,\cite{Landry:2011,Granucci:2010,Jaeger:2012} can resolve these issues and give accurate charge transfer dynamics upon carefully chosen schemes.\cite{Subotnik:2016a} Other recent simulations of PI-PCET based on numerical exact methods such as hierarchical equations of motion (HEOM),\cite{Shi:2017} matching-pursuit/split-operator Fourier transform (MP/SOFT),\cite{Kim:2009} or {\it ab initio} multiple spawning (AIMS)\cite{Pijeau:2017} can provide accurate time-dependent reaction dynamics. However, the numerical costs of these calculations will limit their scope of applications in simulating complex systems with many electronic states and nuclear DOF. It is thus ideal to use accurate yet efficient trajectory-based quantum dynamics approaches, which are essentially different from traditional MQC methods, to directly simulate PI-CPET reaction dynamics.

In this study, we apply a recently developed quasi-diabatic (QD)\cite{Mandal:2018} scheme to directly propagate PI-PCET quantum dynamics with diabatic Partial Linearized Density Matrix (PLDM) path-integral method.\cite{HuoPLDM} As an example of recently developed accurate diabatic trajectory-based quantum dynamics approaches, PLDM uses a consistent dynamical footing for describing all DOFs. Further, we treat both transferring electron and proton quantum mechanically by describing them in their {\it adiabatic vibronic} states, and directly use these adiabatic states as the {\it quasi-diabatic} states during the QD-PLDM propagation. Our numerical results demonstrate the importance of treating proton quantum mechanically for obtaining accurate PI-CPET dynamics, as well as the crucial role of solvent fluctuations and vibrational relaxations that dictate the reaction pathways. We also demonstrate how various reaction regimes can exhibit different kinetic isotope effects (KIE). The QD scheme outlined in this paper does not require any tedious efforts for building strict diabatic system-bath models,\cite{Ananth:2017,Kretchmer:2013} and can be directly generalized to perform {\it ab-initio} on-the-fly simulations.\cite{Goyal:2015,Goyal:2016} This work opens the possibility for studying PI-PCET reaction in realistic chemical systems through combining accurate diabatic quantum dynamics approaches with efficient adiabatic electronic structure calculations. 

\section{Theory and Method}
\subsection{Partial Linearized Density Matrix Method}\label{sec:pldm}
We provide a brief outline of the Partial Linearized Density Matrix (PLDM) path-integral approach.\cite{HuoPLDM,Huo2016ARPC} We begin with expressing the total Hamiltonian as follows
\begin{equation}\label{eqn:totham1}
\hat H = \hat T + \hat{V}_\mathrm{el}(\hat{\bf r},\hat{\bf R}),
\end{equation}
where $\hat{\bf r}$ and $\hat{\bf R}$ represent the electronic and nuclear coordinate operators, $\hat T$ is the nuclear kinetic energy operator, and $\hat{V}_\mathrm{el}$ represents the ``electronic Hamiltonian". Under a set of {\it strict diabatic} states $\{|i\rangle, |j \rangle\}$ which do not explicitly depend on the nuclear configuration, the total Hamiltonian can be expressed as follows
\begin{equation}
\hat H = \hat T + \sum_{ij}^{N} V_{ij}(\hat{\bf R})|i \rangle\langle j|, 
 \end{equation}
where $V_{ij}(\hat{\bf R})=\langle i|\hat{V}_\mathrm{el}(\hat{\bf r},\hat{\bf R})|j\rangle$ is the state-dependent potential for the electronic Hamiltonian operator, and $N$ is the total number of electronic states.

Using the Meyer-Miller-Stock-Thoss\cite{Meyer:1979,StockThossPRL97,StockThossPRA99} (MMST) mapping representation to transform the discrete electronic states into continuous variables, we have $|i \rangle\langle j | \rightarrow  {\hat a}_{i}^\dagger {\hat a}_{j} $, where ${\hat a}^\dagger_{i} =({\hat q}_{i} - i{\hat p}_{i})/\sqrt{2}$. With this transformation, the non-adiabatic transitions between electronic states are exactly mapped onto the classical motion of fictitious harmonic oscillators.\cite{StockThossPRL97,StockThossPRA99} Thus, MMST mapping Hamiltonian provides a consistent classical footing for both electronic and nuclear DOFs.

Expressing the full density matrix operator with the real-time path-integral expression, then applying a partial linearization approximation\cite{HuoPLDM} selectively to the nuclear DOF and keeping the explicit propagation of the electronic mapping DOF for both forward and backward paths, we arrive at the PLDM expression for computing reduced density matrix\cite{HuoPLDM,Huo2016ARPC}
\begin{align}\label{eq:redmat}
&\rho_{ij}(t)=\mathrm{Tr}_\mathrm{\bf R}\left[\hat{\rho}(0)e^{i\hat{H}t/\hbar}|i\rangle\langle j|e^{-i\hat{H}t/\hbar}\right]\\
&\approx \sum_{kl}\int d{\bf R} \frac{d{\bf P}}{ 2\pi\hbar}d{\bf q} d{\bf p} d{\bf q'} d{\bf p'} G_0G'_0 [\hat{\rho}(0)^\mathrm{W}_{kl}]  T_{ki}(t) T'_{jl}(t),\nonumber
\end{align}
where $T_{ki}(t)=\frac{1}{2}(q_{i}(t)+ip_{i}(t))(q_{k}(0)-ip_{k}(0))$ and $T'_{jl}(t)=\frac{1}{2}(q_{l}(0)+ip_{l}(0))(q_{j}(t)-ip_{j}(t))$ are the electronic transition amplitudes, and $[\hat{\rho}(0)^\mathrm{W}_{kl}]$ is the partial Wigner transform (with respect to the nuclear DOF) of the $kl^\mathrm{th}$ matrix element of the initial total density operators $\hat{\rho}(0)$.  The initial distribution of electronic DOF is $G_0({\bf q},{\bf p})= e^{-\frac{1}{2}\sum_{m} (q_{m}^2 + p_{m}^2 )}$ and  $G'_0({\bf q}',{\bf p}') =  e^{-\frac{1}{2}\sum_{n} ({q'}_{n}^{2} + {p'}_{n}^{2} )}$.

Classical trajectories are used to evaluate the approximate time-dependent reduced density matrix in Eqn.~\ref{eq:redmat}. The forward mapping variables are evolved based on the Hamilton's equations of motion\cite{HuoPLDM,Huo2016ARPC}
\begin{equation}  \label{eq:mapeqn} 
\dot q_{i} = \partial h / \partial p_{i}; ~~~\dot p_{i} = -\partial h / \partial q_{i},
\end{equation}
where $h$ is the {\it classical} mapping Hamiltonian\cite{HuoPLDM,HuoMol} with the following expression
\begin{equation} \label{eq:mapham} 
h({\bf p},{\bf q},{\bf R})={1\over2}\sum_{ij}V_{ij}(R)\left(p_{i}p_{j}+q_{i}q_{j}\right).
\end{equation}
The backward mapping variables are propagated with the similar equations of motion governed by $h({\bf p'},{\bf q'},{\bf R})$. The nuclei are propagated with the PLDM force\cite{HuoPLDM} with the following expression
\begin{equation}\label{force}
{\bf F}_\mathrm{PL}=-{1\over 4}\sum_{ij}\nabla V_{ij}(R)\bigg[p_{i}p_{j}+q_{i}q_{j}+p'_{i}p'_{j}+q'_{i}q'_{j}\bigg].
\end{equation}

PLDM uses consistent dynamical footing for both electronic and nuclear DOFs and thus, accurately describes their coupled motion. In contrast,  widely used mixed quantum-classical methods such as Ehrenfest or FSSH\cite{Tully:1990,TullyNAMD} treat quantum and classical DOFs on different footings, which causes the breakdown of detailed balance\cite{Schmidt:2008} or creating the artificial electronic coherence.\cite{Tully:1990,Subotnik:2016a} In addition, compared to the closely related methods that fully linearize both mapping and nuclear DOFs,\cite{miller2001semiclassical,miller2009semiclassical,kelly2012mapping} PLDM retains full dynamical propagation along both forward and backward paths for the mapping DOF, thus achieving a more accurate description of the electronic dynamics.\cite{HuoPLDM,KapralMQCL,HuoJCP13} PLDM has already been successfully applied to simulate a broad range of non-adiabatic processes, including excitation energy transfer dynamics,\cite{Huo2016ARPC,Huo2015PCCP} electron transfer reactions,\cite{HuoJCP13} singlet fission quantum dynamics,\cite{MariaJPCL} and nonlinear optical spectroscopy calculations.\cite{Provazza:2018}

It is worth mentioning that the numerical cost of PLDM scales as $N^2$, where $N$ is the total number of electronic states. This is similar to recently developed methods, including symmetrical windowing quasi-classical (SQC)\cite{Cotton:2016} and forward-backward trajectory solution to the quantum-classical Liouville equation (QCLE).\cite{KapralMQCL} In contrast, the numerical cost of Redfield theory\cite{Berkelbach:2014} or generalized quantum master equation (GQME)\cite{Kelly:2015} scales as $N^4$, though more accurate results can be obtained under several particular parameter regimes.

\subsection{Quasi-Diabatic Propagation Scheme}
Most of the routinely available electronic structure methods are formulated in the {\it adiabatic} representation. However, a large number of recently developed non-adiabatic dynamics methods,\cite{Cotton:2016,KapralMQCL,Kelly:2015} including PLDM,\cite{HuoPLDM} are formulated in the {\it diabatic} representation. Thus, the typical strategy for applying these new methods to ``real'' molecular systems is to reformulate them in the adiabatic representation,\cite{ananth2007,MillerAdiabatic,HuoJCP2012,Hsieh13mol} which usually requires tedious theoretical efforts. Moreover, the adiabatic version of these methods are computationally inconvenient due to the presence of the first and second order derivative couplings,\cite{MillerAdiabatic} which could potentially lead to numerical instabilities during dynamical propagations. 

To address this discrepancy between the accurate {\it diabatic} quantum dynamics approaches and routinely available electronic structure calculations in the {\it adiabatic} states, we have developed Quasi-Diabatic (QD) propagation scheme,\cite{Mandal:2018} a general approach which allows interfacing {\it adiabatic} electronic structure calculations with {\it diabatic} trajectory-based quantum dynamics methods. Here, we provide a brief summary of this scheme, whereas the details of the algorithm can be found in Ref.~\citenum{Mandal:2018}. 

Consider a short-time propagation of the nuclear DOF during $t\in[t_1, t_2]$, where the nuclear positions evolve from ${\bf R}(t_1)$ to ${\bf R}(t_2)$, with the corresponding adiabatic states $\{|\Phi_{\alpha}({\bf R}(t_1))\rangle\}$ and $\{|\Phi_{\mu}({\bf R}(t_2))\rangle\}$. These adiabatic states are defined as the eigenstates of electronic part of the Hamiltonian
\begin{equation}\label{eqn:adia}
\hat{V}_\mathrm{el}(\hat{\bf r},\hat{\bf R})|\Phi_{\alpha}({\bf R})\rangle=E_{\alpha} ({\bf R})|\Phi_{\alpha}({\bf R})\rangle, 
\end{equation}
where $|\Phi_{\alpha}({\bf R})\rangle$ are the adiabatic states with the corresponding eigenenergies $E_{\alpha} ({\bf R})$, both of which explicitly depend on the nuclear coordinates.

The {\it central idea} of the QD propagation scheme\cite{Mandal:2018}  is to use the nuclear geometry at time $t_1$ as the reference geometry, ${\bf R_{0}}\equiv {\bf R}(t_1)$, and the adiabatic basis $\{|\Phi_{\alpha}({\bf R}(t_1))\rangle\}$ as the {\it quasi-diabatic} basis during this short-time quantum dynamics propagation, such that
\begin{equation}
|\Phi_{\alpha}({\bf R_{0}})\rangle\equiv|\Phi_{\alpha}({\bf R}(t_1))\rangle,~~\mathrm{for}~t\in[t_1,t_2].
\end{equation}

With the above QD basis, the derivative couplings vanish in a trivial way during this short-time propagation, and $\hat{V}_{\text{el}}(\hat{\bf r};{\bf R})$ has off-diagonal elements. We emphasize that there is always a non-removable part of the derivative coupling over the {\it entire configurational space} for polyatomic systems.\cite{Mead:1982}  This is a well-known result in literature.\cite{VanVoorhis,SubotnikDIA} Here, the QD scheme circumvents this challenge by requiring {\it locally-defined diabatic} states, such that the derivative couplings vanish in this {\it configurational subspace} during a given short-time propagation.

Because of the diabatic nature of the QD basis during this short-time propagation, one can use any diabatic based approach to propagate the quantum dynamics. These approaches usually require diabatic energies, electronic couplings, and nuclear gradients, which can be conveniently computed\cite{Mandal:2018} under the QD basis, $\{| \Phi_\alpha({\bf R_0})\rangle\}$. For example, one can easily evaluate the matrix elements $V_{\alpha\beta}({\bf R}(t))  = \langle \Phi_\alpha ({\bf R_0})| \hat V_\mathrm{el} (\hat{\bf r}; {\bf R}(t))|  \Phi_\beta({\bf R_0})\rangle$ at both ${\bf R}(t_1)$ and ${\bf R}(t_2)$ as follows 
\begin{eqnarray}
V_{\alpha\beta}({\bf R}(t_{1})) &=&\langle \Phi_\alpha ({\bf R}_0)| \hat V_\mathrm{el} (\hat{\bf r}; {\bf R}(t_1))| \Phi_\beta({\bf R_0})\rangle\\
V_{\alpha\beta}({\bf R}(t_{2})) &=&\sum_{\mu\nu}  b_{\alpha \mu}\langle \Phi_{\mu}({\bf R}(t_{2}))| \hat {V}_\mathrm{el} (\hat{\bf r};{\bf R}(t_2))|\Phi_{\nu}({\bf R}(t_{2})) \rangle b^{\dagger}_{\beta\nu},\nonumber
\end{eqnarray}
where the first expression is simply equal to $E_{\alpha}({\bf R}(t_1))\delta_{\alpha\beta}$, and the second expression contains the basis transformation matrix elements $b_{\alpha\mu}= \langle \Phi_{\alpha}({\bf R_0})|\Phi_{\mu}({\bf R}(t_{2}))\rangle$ and $b^{\dagger}_{\beta\nu} = \langle \Phi_{\nu}({\bf R}(t_{2}))|\Phi_{\beta}({\bf R_0})\rangle$.  The time-dependent matrix elements $V_{\alpha\beta}({\bf R}(t))$ can then be obtained by a linear interpolation between $V_{\alpha\beta}({\bf R}(t_{1}))$ and $V_{\alpha\beta}({\bf R}(t_{2}))$ as follows\cite{Rossky-Webster}
\begin{equation}\label{eqn:interpolation}
V_{\alpha\beta}({\bf R}(t))= V_{\alpha\beta}(\mathbf{R}(t_{1}))+\frac {(t - t_{1})}{(t_{2} - t_{1})}\bigg[V_{\alpha\beta}(\mathbf{R}(t_{2})) - V_{\alpha\beta}(\mathbf{R}(t_{1}))\bigg].
\end{equation} 
Similarly, the nuclear gradients on electronic Hamiltonian matrix elements $\nabla V_{\alpha\beta}({\bf R}(t_{2}))\equiv \partial V_{\alpha\beta}({\bf R}(t_{2}))/\partial {\bf R}$ are evaluated as\cite{Mandal:2018}
\begin{eqnarray} \label{eqn:nucgrad}
&&\nabla V_{\alpha\beta}({\bf R}(t_{2})) = \nabla \langle \Phi_\alpha({\bf R_{0}})| \hat {V}_\mathrm{el}({\hat{\bf r}};{\bf R}(t_2))|  \Phi_\beta ({\bf R_{0}})\rangle \nonumber \\
&&=\langle \Phi_\alpha ({\bf R_{0}})| \nabla \hat{V}_\mathrm{el} ({\hat{\bf r}};{\bf R}(t_2))| \Phi_\beta ({\bf R_{0}})\rangle \\
&&=\sum_{kl}  b_{\alpha\mu} \langle \Phi_{\mu}({\bf R}(t_{2}))|\nabla \hat{V}_\mathrm{el} (\hat{\bf r};\mathbf{R}(t_2))|\Phi_{\nu}({\bf R}(t_{2}))\rangle b^{\dagger}_{\beta\nu},\nonumber
\end{eqnarray}  
where $\langle \Phi_{\mu}({\bf R}(t_{2}))|\nabla \hat{V}_\mathrm{el} (\hat{\bf r};\mathbf{R}(t_2))|\Phi_{\nu}({\bf R}(t_{2}))\rangle$ is the nuclear gradient evaluated under the adiabatic basis $\{|\Phi_{\mu}({\bf R}(t_{2}))\rangle\}$, which is readily available from most electronic structure methods.

During the next short-time propagation segment $t\in[t_2,t_3]$, we adapt a new reference geometry ${\bf R'_{0}}\equiv {\bf R}(t_2)$ and {\it quasi-diabatic} basis $|\Phi_{\mu}({\bf R'_{0}})\rangle\equiv|\Phi_{\mu}({\bf R}(t_2))\rangle$. With the nuclear geometry close to the reference geometry at every single step, the QD representation remains to be a convenient and compact basis in each short-time propagation segment. Between $[t_1,t_2]$ propagation and $[t_2,t_3]$ propagation segments, all of these quantities will be transformed from $\{|\Phi_{\alpha}({\bf R_{0}})\rangle\}$ to $\{|\Phi_{\mu}({\bf R'_{0}})\rangle\}$ basis. In particular, we use the following expressions to transform the mapping variables from the previous to the current QD basis between every two consecutive propagation steps
\begin{eqnarray}
q_{\mu}&\leftarrow&\sum_{\alpha} q_{\alpha} \langle \Phi_{\alpha}({\bf R}(t_{1}))| \Phi_{\mu}({\bf R}(t_{2}))\rangle \\
p_{\mu}&\leftarrow&\sum_{\alpha} p_{\alpha} \langle \Phi_{\alpha}({\bf R}(t_{1}))| \Phi_{\mu}({\bf R}(t_{2}))\rangle.\nonumber
\end{eqnarray}

Here, we use PLDM outlined in Sec.~\ref{sec:pldm} as the diabatic dynamics method in the QD propagation scheme and refer this approach as {\bf QD-PLDM}.\cite{Mandal:2018} This approach provides the same accuracy for non-adiabatic dynamics as obtained from straight diabatic PLDM, with the additional capability to use adiabatic states and nuclear gradients, obtained from electronic structure calculations, for dynamics propagation.

Previous theoretical work with diabatic quantum dynamics approaches\cite{Ananth:2012,Ananth:2017,Kretchmer:2013} for simulating PCET reactions usually require parametrizing the original model system into a strict diabatic system-bath model.\cite{Ananth:2017,Kretchmer:2013} This parametrization process requires tedious efforts  and remains a highly non-trivial task and significant challenge for atomistic simulations.\cite{Ananth:2017} Here, the QD scheme allows directly propagating quantum dynamics by using {\it adiabatic vibronic} basis with diabatic dynamics approaches, thus explicitly avoid any additional efforts for building strict diabatic models.

Further, for the PI-PCET dynamics investigated in this study, the QD scheme provides an additional advantage of reducing the number of electronic states required for dynamical propagation. In the case where the diabatic electron-proton basis (see section III) are treated with MMST mapping variables, one can directly use diabatic PLDM for the dynamical propagation due to the {\it strict diabatic} nature of this basis. However, the computational cost is significantly increased compared to the QD scheme, due to the large number of the diabatic states ($N\sim 100$ in this work). Such technical challenges can be resolved by using a more ``compact" {\it adiabatic} vibronic basis along a given trajectory, {\it i.e.}, the QD basis, which only requires the few low-lying adiabatic states ($N \sim 10$ in this work) which directly participate in the non-adiabatic transitions. 

Thus, the QD scheme provides a convenient propagation framework and a seamless interface for diabatic quantum dynamics approaches with adiabatic electronic structure calculations.\cite{Mandal:2018}

\section{Details of Model Calculations}
\subsection{Model system} 
The PI-PCET model used in this study is expressed as 
\begin{equation}\label{eqn:totham}
\hat{H}=\hat{H}_\mathrm{ep}+\hat{H}_\mathrm{sb},
\end{equation}
where $\hat{H}_\mathrm{ep}$ describes the electron-proton free-energy surfaces, and $\hat{H}_\mathrm{sb}$ describes the solvent-bath interaction. In this work, we adapt two commonly used models for $\hat{H}_\mathrm{ep}$, with one explicitly contains a collective solvent coordinate (ET coordinate)\cite{Hazra:2010,Hazra:2011} which we refer as Model I, and the other that does not contains a special collective solvent coordinate\cite{Venkataraman:2009,Cotton:2014} which we refer as Model II. Note that these two models are related with each other through a simple coordinate transformation,\cite{Cao97jcp} and are equivalent if the latter one has a brownian spectral density.\cite{Cao97jcp,Garg85jcp}

Here, we provide detailed expressions of Model I, with the parameters for solvent-bath interactions provided in Appendix A. The Hamiltonian of Model II is provided in Appendix B. All of the results presented in this work is based on Model I, except those presented in Fig.~\ref{fig:vib} for benchmarking purpose. 

The electron-proton Hamiltonian $\hat{H}_\mathrm{ep}$ in Eqn.~\ref{eqn:totham} of Model I is expressed as 
\begin{equation}\label{eqn:langeham}
\resizebox{1.\hsize}{!}{
$\hat{H}_\mathrm{ep}=\hat{T}_\mathrm{p}+$
\Bigg[
  \begin{tabular}{cc}
  $U^\mathrm{D}(\hat{r}_\mathrm{p})+\frac{1}{2}M_\mathrm{s}\omega^{2}_\mathrm{s} R_\mathrm{s}^2$ & $V_\mathrm{DA}$\\
   $V_\mathrm{DA}$ & $U^\mathrm{A}(\hat{r}_\mathrm{p})+\frac{1}{2}M_\mathrm{s}\omega^{2}_\mathrm{s} (R_\mathrm{s}- R_\mathrm{s}^0)^2-\Delta$\\
  \end{tabular}
  \Bigg],}
\end{equation}
where the first term $\hat{T}_\mathrm{p}$ represents the kinetic energy operator of the proton, and the second term represents the electron-proton interaction potential in the diabatic donor $|\mathrm{D} \rangle$ and acceptor $|\mathrm{A} \rangle$ electronic states, with $V_\mathrm{DA}=0.03$ eV as the coupling between the two electronic states. Here, $\hat{r}_\mathrm{p}$ is the proton coordinate, and $U^\mathrm{D}(\hat{r}_\mathrm{p})$ and $U^\mathrm{A}(\hat{r}_\mathrm{p})$ represent the proton free-energy profile associated with $|\mathrm{D}\rangle$ and $|\mathrm{A}\rangle$ states, with the following expressions
\begin{equation}\label{eqn:donor-accep}
U^\mathrm{D}(\hat{r}_\mathrm{p})=\frac{1}{2}m_{p}\omega_\mathrm{p}^{2} (\hat{r}_\mathrm{p}-r_\mathrm{p}^\mathrm{D})^{2};~~U^\mathrm{A}(\hat{r}_\mathrm{p})=\frac{1}{2}m_{p}\omega_\mathrm{p}^{2} (\hat{r}_\mathrm{p}-r_\mathrm{p}^\mathrm{A})^{2}.
\end{equation}
In this work, we use $r_\mathrm{p}^\mathrm{D}=0$ and $r_\mathrm{p}^\mathrm{A}=0.5$ \AA~as the minima of proton free-energy profile associated with the electronic donor and acceptor states. $m_\mathrm{p}=1.0073$ amu and $\omega_\mathrm{p}=3000~\mathrm{cm}^{-1}$ are the mass and vibrational frequency of the proton. $R_\mathrm{s}$ represents the collective solvent coordinate for electron transfer where $M_\mathrm{s}$ and $\omega_\mathrm{s}=\sqrt{f_0/M_\mathrm{s}}$ are the mass and frequency of this coordinate with $f_0$ as the force constant. $\lambda$ is the solvent reorganization energy, and $\Delta$ is the driving force (bias) of the reaction. In this model, the proton and the solvent DOF do not explicitly interact with each other; rather, $\hat{r}_\mathrm{p}$ directly interacts with various electronic states, which in turn interact with the solvent. In addition, we choose three possible driving forces,\cite{Hazra:2010} with $\Delta=0$ ({\bf Model IA}), $\Delta=1$ eV ({\bf Model IB}), and $\Delta=3.51$ eV ({\bf Model IC}). All the other parameters are provided in Appendix A.

\begin{figure}[t]
\centering
\includegraphics[width=0.9\linewidth]{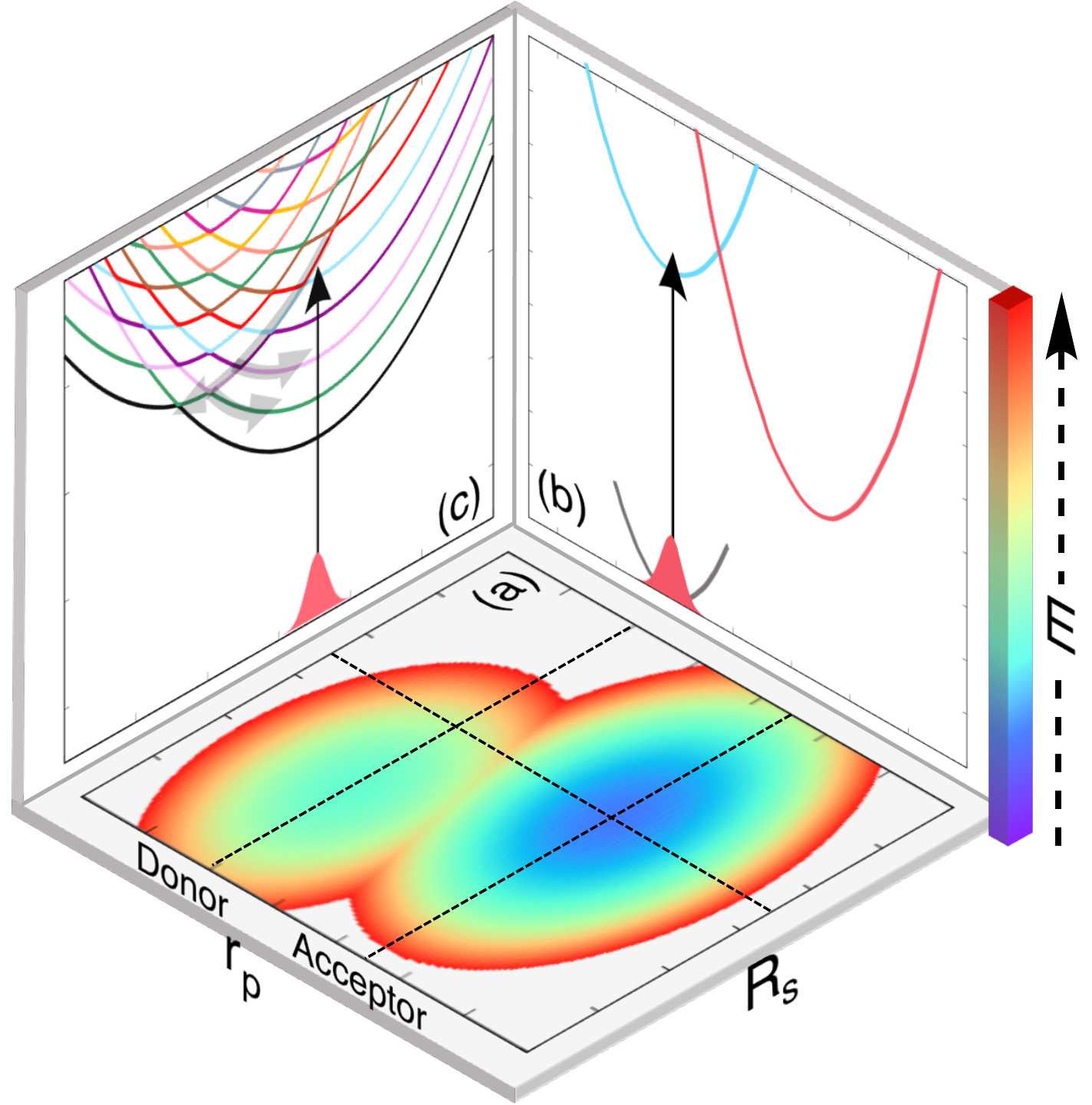}
\caption{Schematic illustration of the model PI-PCET systems with the driving force $\Delta = 1$ eV.
(a) The adiabatic electron-proton free energy surface of the $\mathrm{S_{1}}$ (first excited electronic state) as a function of proton ($r_\mathrm{p}$) and solvent ($R_\mathrm{s}$) coordinates. (b) The proton free energy diabatic potentials that correspond to the electronic ground state prior to photoexcitation (gray), photoexcited $|\mathrm{D}\rangle$ state (blue), and $|\mathrm{A}\rangle$ state (red) as functions of $r_\mathrm{p}$. (c) Adiabatic electron-proton vibronic free energy surfaces as functions of $R_\mathrm{s}$. 
}
\label{fig:model}
\end{figure}

Fig.~\ref{fig:model} illustrates the electron-proton potential $\hat{H}_\mathrm{ep}$, with the driving force $\Delta= 1$ eV (Model IB). Fig.~\ref{fig:model}a presents the electronic adiabatic free energy surface of the first excited electronic state $\mathrm{S_1}$, as a function of proton coordinate $\hat{r}_\mathrm{p}$ and solvent coordinate $R_\mathrm{s}$. Two dashed lines perpendicular to $\hat{r}_\mathrm{p}$ coordinate indicate the minima for proton donor ($U^\mathrm{D}(\hat{r}_\mathrm{p})$) and acceptor ($U^\mathrm{A}(r_\mathrm{p})$) free energy diabats. One dashed line perpendicular to $R_\mathrm{s}$ indicate the center of the initial solvent distribution. Fig.~\ref{fig:model}b presents the photoexcited electronic state $|\mathrm{D}\rangle$ (blue) and $|\mathrm{A}\rangle$ (red), which correspond to the diagonal elements of $\hat{H}_\mathrm{ep}$ (Eqn.~\ref{eqn:langeham}) evaluated at $R^0_\mathrm{s} =\sqrt{2\lambda/f_0}$ (dashed line perpendicular to $R_\mathrm{s}$ in panel (a)). Black arrow indicates the photoexcitation process of the proton from its harmonic ground state, $U^0=\frac{1}{2}m_\mathrm{p}\omega_\mathrm{p}^{2} \hat{r}_\mathrm{p}^{2}$. Fig.~\ref{fig:model}c depicts the adiabatic electron-proton vibronic free energy surfaces as functions of $R_\mathrm{s}$, which are obtained by diagonalizing $\hat{H}_\mathrm{ep}$ in the vibronic basis (with details described in section~\ref{sec:adia}). In this panel, black arrow indicates the photoexcitation that promotes the initial solvent distribution centered at $\sqrt{2\lambda/f_0}$, whereas gray arrows indicate the subsequent vibrational relaxation process.

The solvent-bath Hamiltonian $\hat{H}_\mathrm{sb}$ in Eqn.~\ref{eqn:totham} is expressed as follows
\begin{equation}
\hat{H}_\mathrm{sb}=\frac{P^{2}_\mathrm{s}}{2M_\mathrm{s}}+\sum_{k}\left[\frac{P_{k}^2}{2M_{k}} + \frac{1}{2}M_{k} \omega_{k}^{2} \left(R_{k}- \frac{c_{k}R_\mathrm{s}}{M_{k} \omega_{k}^{2}}\right)^2\right].
\end{equation}

In the above equation, $R_k$ represents the $k^\mathrm{th}$ bath mode, with the corresponding coupling constant $c_k$ and frequency $\omega_k$ sampled from the following spectral density 
\begin{equation}
J(\omega)={\pi\over2}\sum_{k}{{c^{2}_{k}}\over{M_{k}\omega_k}}\delta(\omega-\omega_k)=f_0\tau_{\text{L}}\omega e^{-{{\omega}\over{\omega_\mathrm{c}}}}.
\end{equation}
Here, $\tau_{\text{L}}$ is the solvent response time (see Appendix A), $M_{k}$ is the mass of the $k^\mathrm{th}$ bath mode, and $\omega_\mathrm{c}$ is the characteristic frequency of the bath that is much faster than the motion of $R_\mathrm{s}$. Here, we choose $\omega_\mathrm{c}=10\omega_\mathrm{s}$ and $M_{k}=M_\mathrm{s}$ for all $k$. 

One can thus perform QD-PLDM simulation with the above total Hamiltonian. Alternatively, we can perform the following equivalent Langevin dynamics\cite{Hazra:2010} that treats the bath implicitly, with the equation of motion for the collective solvent coordinate $R_\mathrm{s}$ as follows
\begin{equation}\label{eqn:langevin}
M_\mathrm{s} \ddot{R_\mathrm{s}}={\bf F}_\mathrm{ep}(R_\mathrm{s})-f_0 \tau_{\text{L}}\dot{R_\mathrm{s}}+{\bf F}_\mathrm{r}(t). 
\end{equation}
In the above Langevin equation, ${\bf F}_\mathrm{ep}(R_\mathrm{s})$ is the PLDM force (see Eqn.~\ref{force}) associated with $\hat{H}_\mathrm{ep}$, the friction force is $-f_0 \tau_{\text{L}}\dot{R_\mathrm{s}}$ with the friction constant $f_0 \tau_{\text{L}}$, and ${\bf F}_\mathrm{r}(t)$ is the random force bounded by the fluctuation-dissipation theorem through equation $\langle {\bf F}_\mathrm{r}(t){\bf F}_\mathrm{r}(0)\rangle=2k_\mathrm{B}Tf_0 \tau_{\text{L}}\delta(t)$.  Here, $\textbf{F}_{\text{r}}(t)$ is modeled as a Gaussian random force with the distribution width\cite{Tully:1979} $\sigma =\sqrt{2k_{\text{B}}T f_0 \tau_{\text{L}}/dt}$, where $k_{\text{B}}$ is the Boltzmann constant and $dt$ is the nuclear time step. The details for generating $\tau_\mathrm{L}$ for a given solvent is provided in Appendix A.

As a consistency check, we have verified that equivalent results (for time-dependent electron-proton reduced density matrix) are obtained with either explicit bath (dynamics with the full Hamiltonian $\hat{H}_\mathrm{el}+\hat{H}_\mathrm{sb}$) or implicit bath (Langevin dynamics in Eqn.~\ref{eqn:langevin}) approach. The equivalency of both approaches have also been recently explored in the condensed-phase ET dynamics\cite{HuoJCP13, Landry:2011,Subotnik:2013,Schwerdtfeger2014} as well as PI-PCET dynamics.\cite{Auer2012} We should also note that when the bath DOF has high vibrational frequency such that $\hbar\omega_\mathrm{s}\gg k_\mathrm{B}T$, linearization approximation\cite{Huo10CP} or classical treatment for the bath can become less accurate.\cite{Wang13JCP, Huo10CP} Thus, performing implicit Langevin dynamics for the bath can provide more accurate results, especially for those approximate quantum dynamics approaches.\cite{Wang13JCP} 

\subsection{Adiabatic Vibronic Surfaces}\label{sec:adia}
In this work, we treat both electron and proton quantum mechanically with the corresponding vibronic states. Thus, the ``electronic part" of the Hamiltonian, {\it i.e.},  $\hat{V}_\mathrm{el}$ in Eqn.~\ref{eqn:totham1} and Eqn.~\ref{eqn:adia}, is defined as $\hat{H}_\mathrm{ep}$ in Eqn.~\ref{eqn:totham}, such that
\begin{equation}\label{eqn:epham}
\hat{V}_\mathrm{el}\equiv \hat{H}_\mathrm{ep}(\hat{T}_\mathrm{p},\hat{r}_\mathrm{p}, \hat{r}_\mathrm{e},R_\mathrm{s}), 
\end{equation}
Thus,  $\hat{V}_\mathrm{el}$ includes proton kinetic energy, electronic potential, electron-proton and electron-solvent interactions. 

In order to obtain the adiabatic vibronic states $|\Phi_{\alpha}({R}_\mathrm{s})\rangle$ for the coupled electron-proton Hamiltonian $\hat{H}_\mathrm{ep}$, we express $|\Phi_{\alpha}({R}_\mathrm{s})\rangle$ with a set of two-particle basis functions as follows
\begin{equation}\label{wfexp}
|\Phi_{\alpha}({R}_\mathrm{s})\rangle=\sum\limits_{i,m} c_{im}^{\alpha}({R}_\mathrm{s})|\phi_\mathrm{e}^i\rangle|\phi_\mathrm{p}^m\rangle, 
\end{equation}
where  $|\phi_\mathrm{e}^i\rangle \in \{|\mathrm{D}\rangle,|\mathrm{A}\rangle\}$ and $|\phi_\mathrm{p}^{m}\rangle$ is chosen to be the $m^\mathrm{th}$ eigenfunction of a quantum harmonic oscillator, with the total Hamiltonian $\hat{H}=\hat{T}_\mathrm{p}+{1\over 2}m_\mathrm{p}\omega_ \mathrm{p}^{2}\hat{r}_\mathrm{p}^2$. Thus, by using $M$ harmonic basis functions for proton and two basis states for electron, the total number of vibronic basis is $N=2M$, and $\hat{H}_\mathrm{ep}$ contains $2M\times 2M$ Hamiltonian matrix $\langle\phi_\mathrm{p}^n|\langle\phi_\mathrm{e}^j|\hat{H}_\mathrm{ep}|\phi_\mathrm{e}^i\rangle|\phi_\mathrm{p}^m\rangle$ under this representation. Because both $U^\mathrm{D}(\hat{r}_\mathrm{p})$ and $U^\mathrm{A}(\hat{r}_\mathrm{p})$ are just simple displaced harmonic oscillator potentials, the matrix elements of $\hat{H}_\mathrm{ep}$ can be obtained analytically by recognizing the basic property of harmonic oscillator as follows 
\begin{eqnarray}
&&\langle \phi_\mathrm{p}^n|\hat{T}_\mathrm{p}+{1\over 2}m_\mathrm{p}\omega_\mathrm{p}^{2}\hat{r}_\mathrm{p}^2|\phi_\mathrm{p}^m\rangle=\bigg(n+{1\over2}\bigg)\hbar\omega_\mathrm{p}\delta_{nm}\\
&&\langle \phi_\mathrm{p}^n|\hat{r}_\mathrm{p}|\phi_\mathrm{p}^m\rangle = \sqrt{\frac{\hbar}{m_\mathrm{p}\omega_\mathrm{p}}}\frac{1}{\sqrt{2}} \bigg(\sqrt{m}\, \delta_{n,m-1}+\sqrt{m+1}\, \delta_{n,m+1}\bigg).\nonumber
\end{eqnarray}

The eigenvalues and the eigenvectors ({\it adiabatic vibronic} basis) are then obtained through direct diagonalization of the $\hat{H}_\mathrm{ep}$ matrix under the above two-particle basis. 

\subsection{Quantum dynamics propagation approaches} 
We perform PLDM quantum dynamics simulations with three different choices for treating the transferring electron and proton, which are summarized as follows 
\begin{itemize}
\item {\bf QD-PLDM}: describe the electron-proton \textit{adiabatic vibronic} basis $|\Phi_{\alpha}({R}_\mathrm{s})\rangle$ as the quasi-diabatic ({\bf QD}) states with MMST mapping variables; propagate the dynamics with QD-PLDM approach,\cite{Mandal:2018}
\item {\bf vib-PLDM}: describe the electron-proton {\it diabatic vibronic} ({\bf vib}) basis $|\phi_\mathrm{e}^i\rangle|\phi_\mathrm{p}^m\rangle$ with MMST mapping variables; propagate the dynamics with straight {\it diabatic} PLDM approach,
\item {\bf el-PLDM}: only describe the electronic ({\bf el}) state  $|\phi_\mathrm{e}^i\rangle$ with MMST mapping variables, whereas the proton is treated through linearization approximation (which gives rise to classical equation of motion and a Wigner initial distribution); propagate the dynamics with straight {\it diabatic} PLDM approach. 
\end{itemize}

Here, we briefly comment on the numerical cost of each approach. The convergence of results in vib-PLDM and QD-PLDM is obtained with a considerably less number of trajectories ($\sim$ 10 times fewer) than in el-PLDM. This is because el-PLDM requires sampling of the Wigner distribution of both the solvent and the proton DOF (with a broad width of the distribution due to large $\omega_\mathrm{p}$), while in vib-PLDM and QD-PLDM the initial sampling is only required for the solvent DOF. 

In addition, the number of states required to be explicitly propagated are different in each one of these schemes. Here, el-PLDM only requires explicit propagation of two states, $|\mathrm{D}\rangle$ and $|\mathrm{A}\rangle$;  vib-PLDM requires explicit propagation of 60-120 diabatic vibronic basis $\{|\phi_\mathrm{e}^i\rangle|\phi_\mathrm{p}^m\rangle\}$; QD-PLDM only requires 5-20 adiabatic vibronic states $\{|\Phi_{\alpha}({R}_\mathrm{s})\rangle\}$, due to the compactness of these adiabatic states for describing the changing wavefunction. Recall that the numerical cost of PLDM propagation scales as $N^2$ (with $N$ as the total number of states). Thus, the ultimate numerical cost for vib-PLDM is much larger than both QD-PLDM and el-PLDM. QD-PLDM, on the other hand, still requires explicitly diagonalizing $\hat{H}_\mathrm{ep}$ matrix with a numerical cost of $N^3$, besides the cost for dynamical propagation.

\subsection{Simulation details}
Here, we provide the simulation details for Model I, whereas the corresponding details for Model II are provided in Appendix B. 

The converged results for Model I are obtained with 2400 trajectories for QD-PLDM or vib-PLDM propagations, with a time step of $dt=0.024$ fs (1 a.u.). The total number of vibrational basis $\{|\phi_\mathrm{p}^{m}\rangle\}$ used in Model IA is 30, {\it i.e.}, $m=0,1,...29$ for the Harmonic oscillator eigenstates. The total number of vibrational basis used in Model IB and IC are 40 and 60, respectively. We have carefully checked the convergence of our results with additional 10 vibrational basis, which generates numerically identical results. For QD-PLDM propagation, we only used the first 5, 10, and 20 time-dependent low-lying adiabatic vibronic states as the QD states for the electron-proton description. We have also carefully checked the convergence of the QD propagation scheme with additional 10 more QD basis, which also provides the identical results.

In all calculations with Model I, the system is initially prepared in the proton vibrational ground state $|\phi_\mathrm{p}^{0}\rangle$ of the electronic ground state $|\mathrm{S_0}\rangle$. The system is then excited to the $|\mathrm{D}\rangle$ state through Franck-Condon process, which generate the initial state described by the following total density operator 
\begin{equation}
\hat{\rho}(0)=|\Phi(0)\rangle \langle\Phi(0)|\otimes \hat{\rho}_\mathrm{s}.
\end{equation}
Here, the initial electron-proton quantum state is expressed as 
\begin{equation}
|\Phi(0)\rangle=|\mathrm{D}\rangle|\phi_\mathrm{p}^{0}\rangle, 
\end{equation}
and $\hat{\rho}_\mathrm{s}$ is the density operator of the solvent. PLDM requires the partial Wigner transform of the total density operator $\hat{\rho}(0)$ which can be easily obtained (due to the its simple direct product form) as follows
\begin{equation}
[\hat{\rho}(0)^\mathrm{W}]=|\Phi(0)\rangle \langle\Phi(0)|\otimes \rho_\mathrm{s}^\mathrm{W},
\end{equation} 
with the following Wigner density for the solvent 
\begin{equation}
\rho_\mathrm{s}^\mathrm{W} = { \omega_\mathrm{s} \Gamma_\mathrm{s}} e^{-\Gamma_\mathrm{s} \big[{{P_\mathrm{s}^2\over {2M_\mathrm{s}}} + {1\over2}M_\mathrm{s}\omega_\mathrm{s}^2({R_\mathrm{s}}-{R_\mathrm{s}^0})^2\big]}}.
\end{equation}
Here,  $\Gamma_\mathrm{s}= (2/\omega_\mathrm{s})\tanh(\omega_\mathrm{s}/2k_\mathrm{B}T)$ and $\omega_\mathrm{s} = \sqrt{f_0/M_\mathrm{s}}$. In this study, we choose $R^0_\mathrm{s} =\sqrt{2\lambda/f_0}$ that corresponds to the minimum of the proton acceptor free energy diabatic surface. Further, we use the focused initial conditions\cite{FocusedBonella} to facilitate the convergence of the sampling for the mapping variables, which obey the distribution governed by $G_0({\bf p,q})$ and $G'_0 ({\bf p',q'})$ (see Sec.\ref{sec:pldm} for details). In vib-PLDM, this means that $q_{\xi}=q'_{\xi}=\delta_{\xi\eta}$ and $p_{\xi}=-p'_{\xi}=\delta_{\xi\eta}$ where $|\eta\rangle=|\Phi(0)\rangle=|\mathrm{D}\rangle|\phi_\mathrm{p}^{0}\rangle$ and $|\xi\rangle=|\phi^{i}_\mathrm{e}\rangle|\phi^{m}_\mathrm{p}\rangle$. Whereas in QD-PLDM, the initial values of the corresponding mapping variables are obtained through the following expressions
\begin{equation}
q_{\alpha}=\sum_{im} q_{\xi} c^{\alpha}_{im};~~p_{\alpha}=\sum_{im} p_{\xi} c^{\alpha}_{im},
\end{equation}
where $c^{\alpha}_{im}$ is the eigenfunction coefficients obtained from Eqn.~\ref{wfexp} and $|\xi\rangle=|\phi^{i}_\mathrm{e}\rangle|\phi^{m}_\mathrm{p}\rangle$.

\section{Results and Discussions}
\begin{figure}[t]
\centering
\includegraphics[width=0.9\linewidth]{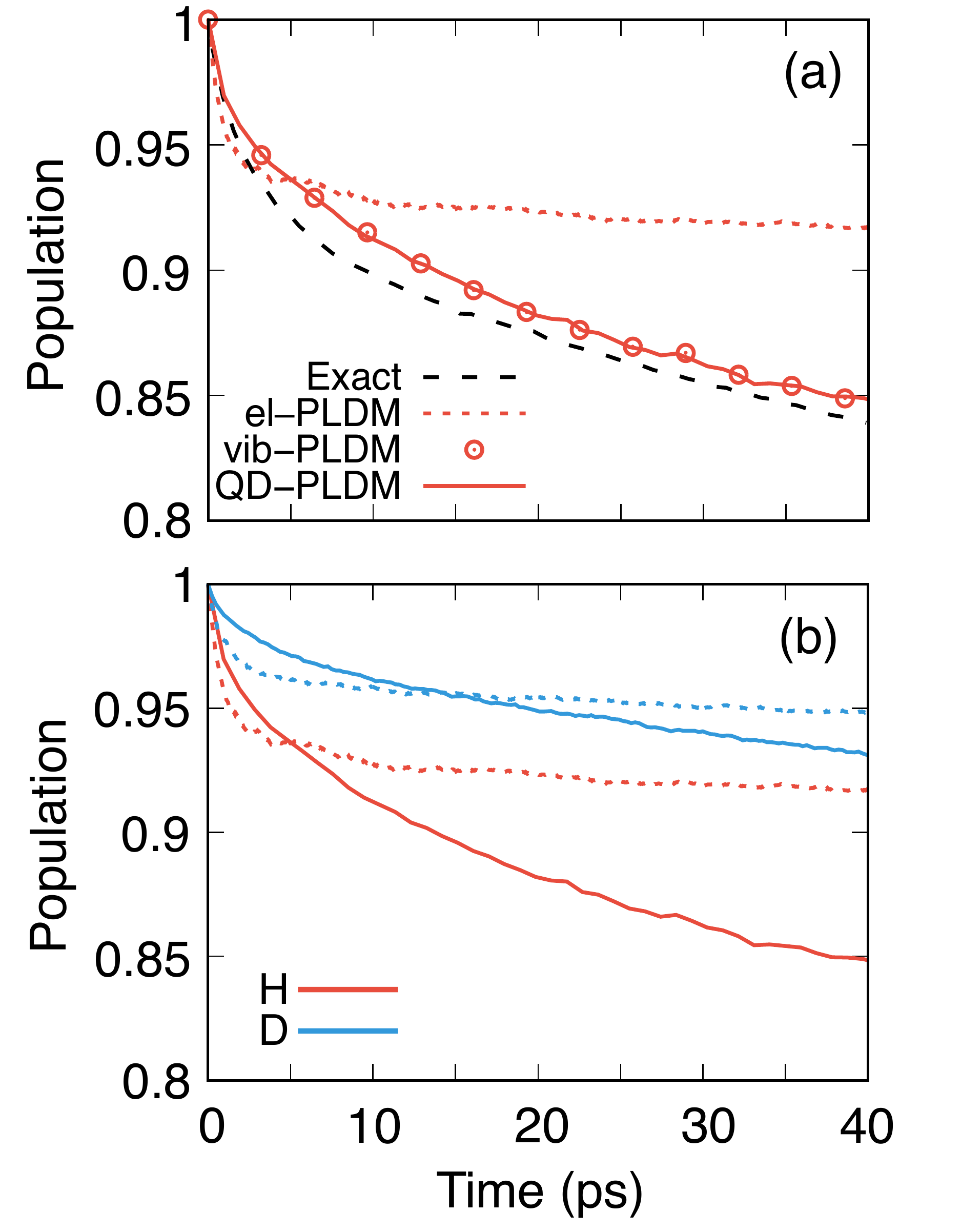}
\caption{(a) Diabatic population of the photoexcited donor state obtained from exact calculations (black dashed line), el-PLDM (dotted line), vib-PLDM  (open circles) and QD-PLDM (solid line). (b) Comparison of donor diabatic state population for proton (red) and deuterium (blue) with el-PLDM (dashed line) and vib-PLDM (solid line).}
\label{fig:vib}
\end{figure}

Fig.~\ref{fig:vib} presents the {\it diabatic electronic}  population of the $|\text{D}\rangle$ state obtained from el-PLDM (dotted line), vib-PLDM (open circle), and QD-PLDM (solid line) for Model II. In addition, in order to assess the accuracy of these approaches, we present results obtained from reduced density matrix (RDM) formalism\cite{Venkataraman:2009,Cotton:2014} which provides the exact results (black dashed line) under the weak system-bath coupling regime for Model II. It can be clearly seen from Fig.~\ref{fig:vib}a that el-PLDM which propagates the motion of proton classically (from its initial Wigner distribution), does not provide accurate electronic dynamics and causes large deviation from the exact result at a longer time. Similar deviation has also been reported by using symmetrical quasi-classical (SQC) method\cite{Cotton:2014} with the same classical treatment for proton. On the other hand, quantizing the proton with vibronic basis in both vib-PLDM and QD-PLDM approaches leads to accurate PI-PCET dynamics. 

We would like to mention that treating vibrational diabatic states with MMST mapping variables can certainly improve the accuracy of the dynamics, as been demonstrated with other recent theoretical works. In one example, vib-PLDM is used to compute 2-dimensional electronic spectra\cite{Provazza:2018} in a Frenkel excitons model with a high-frequency vibrational mode. in another example, a new method, so called extended SQC approach\cite{Kananenka:2018} is developed based on similar strategy of vib-PLDM, which provides accurate non-adiabatic dynamics when a highly non-harmonic mode is explicitly quantized with its vibrational eigenbasis. In the third example, the lowest four electron-proton diabatic states are explicitly used in PCET quantum dynamics studies.\cite{Ananth:2017} However, we want to emphasize that in general, vib-PLDM does require a large number of strict diabatic states to be explicitly propagated for investigating PI-PCET reaction. Further, strict diabatic states cannot be easily obtained for real systems, in addition to the computational disadvantage associated with propagating a large number of states. This feature will ultimately limit the scope and applicability of vib-PLDM for investigating PI-PCET. The QD scheme, on the other hand, only requires a small set of adiabatic vibronic states for time-dependent propagation, and thus provides an accurate and efficient theoretical framework for investigating PI-PCET dynamics.

Fig.~\ref{fig:vib}b presents the kinetic isotope effect (KIE) for proton (red) and deuterium (blue) in terms of the donor population. For clarity, here we only present the results obtained from el-PLDM (dotted lines) and QD-PLDM (solid lines), whereas the results from the latter one are identical to vib-PLDM, and agree very well with exact results obtained from RDM\cite{Venkataraman:2009} (not shown here). Because deuterium is essentially a classical particle, the el-PLDM and QD-PLDM approaches provide similar results (blue curves) despite some small deviations. This suggests that treating deuterium classically provide a reasonably accurate dynamics. whereas the quantum nature of the proton requires an explicit propagation with electron-proton vibronic states for accurate results.


\begin{figure*}[t]
\centering
\includegraphics[width=0.9\textwidth]{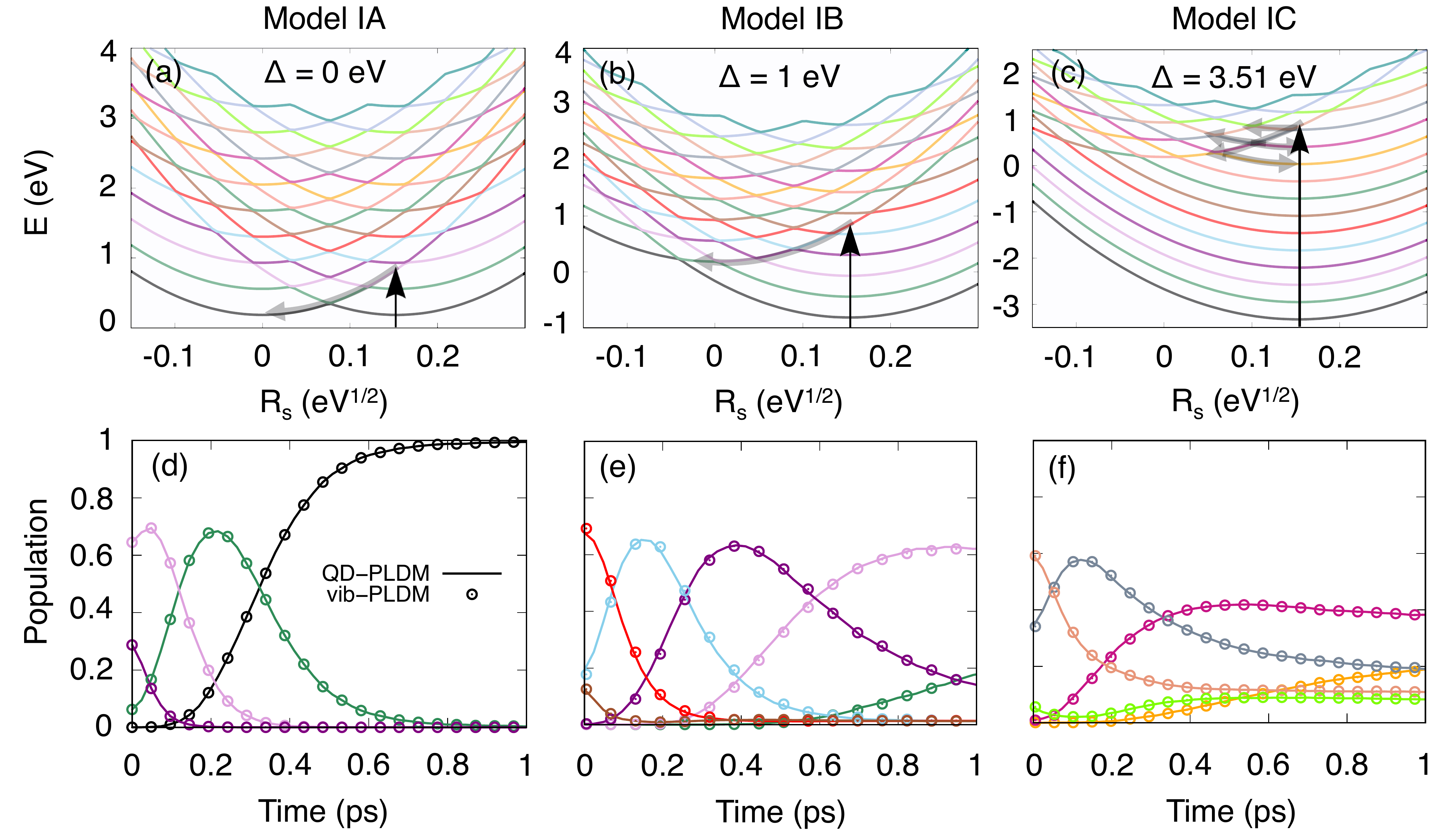}
\caption{(a)-(c) Adiabatic vibronic free energy surfaces as functions of the collective solvent coordinate for Model IA ($\Delta=0$ eV), Model IB ($\Delta=1$ eV), and Model IC ($\Delta=3.51$ eV), with (d)-(f) representing the corresponding adiabatic vibronic populations obtained from vib-PLDM (open circles) and QD-PLDM (solid lines).}
\label{fig:pop}
\end{figure*}

Fig.~\ref{fig:pop} presents the {\it adiabatic vibronic} population of the PI-PCET dynamics, with the adiabatic surfaces of models IA-IC provided in (a)-(c). The initial photoexcitation is illustrated with black arrows, and the subsequent vibrational relaxation pathways are indicated with gray arrows. The corresponding color-coded vibronic state populations are presented in panels (d)-(e), calculated using vib-PLDM (open circles) and QD-PLDM (solid lines) propagations, which are identical. Quantitatively similar results for these vibrational relaxation dynamics have also been obtained from FSSH\cite{Hazra:2010} approach (results not shown), suggesting that the decoherence correction might not have large impact on the short-time vibronic dynamics ($\sim$ 1 ps). 

In Fig.~\ref{fig:pop}a-c, the initial photoexcitation leads to populating a set of high-lying vibronic excited states, followed by vibrational relaxation process that propagates the vibronic wavepacket into low-lying states through non-adiabatic transitions. The grey arrows indicate these vibrational relaxation dynamics during the first 1 ps timescale of the simulations. Compared to the symmetric case in Model IA ($\Delta=0$), the non-zero energy bias in Models IB and IC leads to populating much higher vibronic states at the beginning of the reaction, as well as stabilizing the acceptor state over the donor state that impacts the longer time dynamics. For model IC, the solvent coordinates $R_\mathrm{s}$ directly relaxes to the electronic acceptor side during the first 1 ps (results not shown), whereas in Model IA and IB, $R_\mathrm{s}$ relaxes back to the electronic donor side, suggesting a much slower ET dynamics associated with IA and IB, as will be demonstrated in the next figure. 

Fig.~\ref{fig:KIE} presents the KIE with electronic population dynamics of Model IB and IC. The corresponding dynamics in Model IA is much slower compared to IB and IC, and thus is not shown here. Fig.~\ref{fig:KIE}a depicts the population decay of the donor state for proton (red) and deuterium (blue) in Model IB. It can be seen that there is a clear separation of time scale during the PI-PCET dynamics, with an initial fast vibrational relaxation process during the first 1 ps (that corresponds to results shown in Fig.~\ref{fig:pop}d), followed by a second stage, much slower non-adiabatic dynamics that transfer electronic population from the donor to the acceptor state. In contrast, in the system with deuterium, the donor electronic population does not significantly transfer during the same time scales. These results indicate a large KIE that can be observed in Model IB.

Fig.~\ref{fig:KIE}b presents the same population dynamics for Model IC. One can also observe a similar two-stage dynamical process, with an ultrafast sub-picosecond relaxation process and a relatively slower ($\sim 5$ ps) charge population transfer dynamics. In this model system, KIE is negligible compared to the previous model, whereas the early stage relaxation process for deuterium is even faster than proton. Similar negligible or even slightly inverse KIE\cite{Shi:2017,Goyal:2017,Hazra:2011,Hazra:2010} has also been observed through recent theoretical investigations. The inverse KIE can be easily understood as follows. When tunneling effects are less important, the vibrational relaxation dictates the dynamics,\cite{Goyal:2017,Goyal:2016} and with a larger nuclear mass, deuterium relaxes even faster than proton\cite{Shi:2017} because the vibrational states are closer in energy. Quantum mechanically, the vibrational gap of deuterium is much smaller compared to proton. With the same initial photoexcitation, more high-lying excited vibronic states can be populated for deuterium, and thus promotes the PCET process.\cite{Shi:2017} Further, as been previously discussed,\cite{Hazra:2011,Goyal:2017,Shi:2017} the lack of KIE in the initial stage of PI-PCET cannot exclude the possibility of the concerted transfer of both proton and electron. 

In order to understand the distinctly different KIE in the above two model systems, we compute the time-dependent probability density of the transferring proton/deuterium associated with the donor electronic states $|\Phi_\mathrm{D}(r)|^2=\sum_{\eta\xi}\rho_{\eta\xi}(t) \langle r|\phi_\mathrm{p}^{m}\rangle \langle \phi_\mathrm{p}^{n}|r\rangle$, where $\rho_{\eta\xi}(t)$ is the reduced density matrix in the electron-proton diabatic vibronic basis $\{|\phi^{i}_\mathrm{e}\rangle|\phi^{m}_\mathrm{p}\rangle,|\phi^{j}_\mathrm{e}\rangle|\phi^{n}_\mathrm{p}\rangle\}$, $|\eta\rangle=|\mathrm{D}\rangle|\phi_\mathrm{p}^m\rangle$, and $|\xi\rangle=|\mathrm{D}\rangle|\phi_\mathrm{p}^n\rangle$. Similar expression is used for computing $|\Phi_\mathrm{A}(r)|^2$, where $|\eta\rangle=|\mathrm{A}\rangle|\phi_\mathrm{p}^m\rangle$, and $|\xi\rangle=|\mathrm{A}\rangle|\phi_\mathrm{p}^n\rangle$.

Fig.~\ref{fig:KIE}c-e presents these time-dependent probability densities for the transferring proton/deuterium described above. In Fig.~\ref{fig:KIE}c, there is a significant transfer of the proton probability distribution from the donor to the acceptor states in Model IB, whereas there is no transfer for deuterium probability distribution presented in Fig.~\ref{fig:KIE}e. This different behavior suggests that tunneling of the proton between the donor and acceptor states dominate the PI-PCET dynamics at a longer time, after the initial short-time vibrational relaxation. In contrast, for Model IC, the probability densities for both proton (Fig.~\ref{fig:KIE}d) and deuterium (Fig.~\ref{fig:KIE}f) exhibit very similar time-dependent behavior, suggesting a predominant role of vibrational relaxation and less important role of tunneling in the PI-PCET dynamics.
\begin{figure}[t]
\centering
\includegraphics[width=\columnwidth]{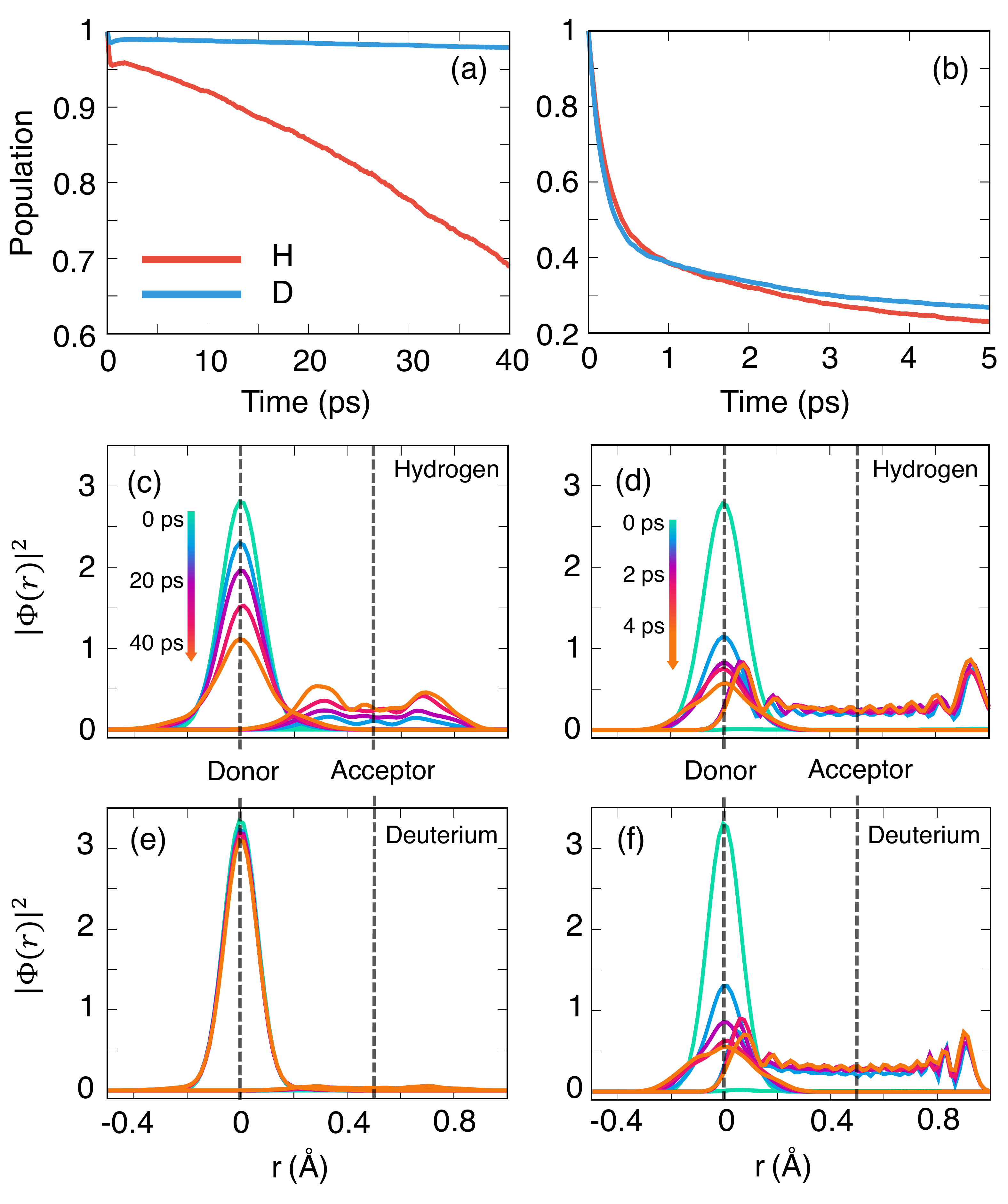}
\caption{Diabatic population of the donor state with proton (red) and deuterium (blue) in Model IB (a) and Model IC (b). The time-dependent probability density of the transferring proton are presented in (c) and (d), while the probability density of the transferring deuterium are presented in (e) and (f).}
\label{fig:KIE}
\end{figure}

\begin{figure}[t]
\centering
\includegraphics[width=0.8\columnwidth]{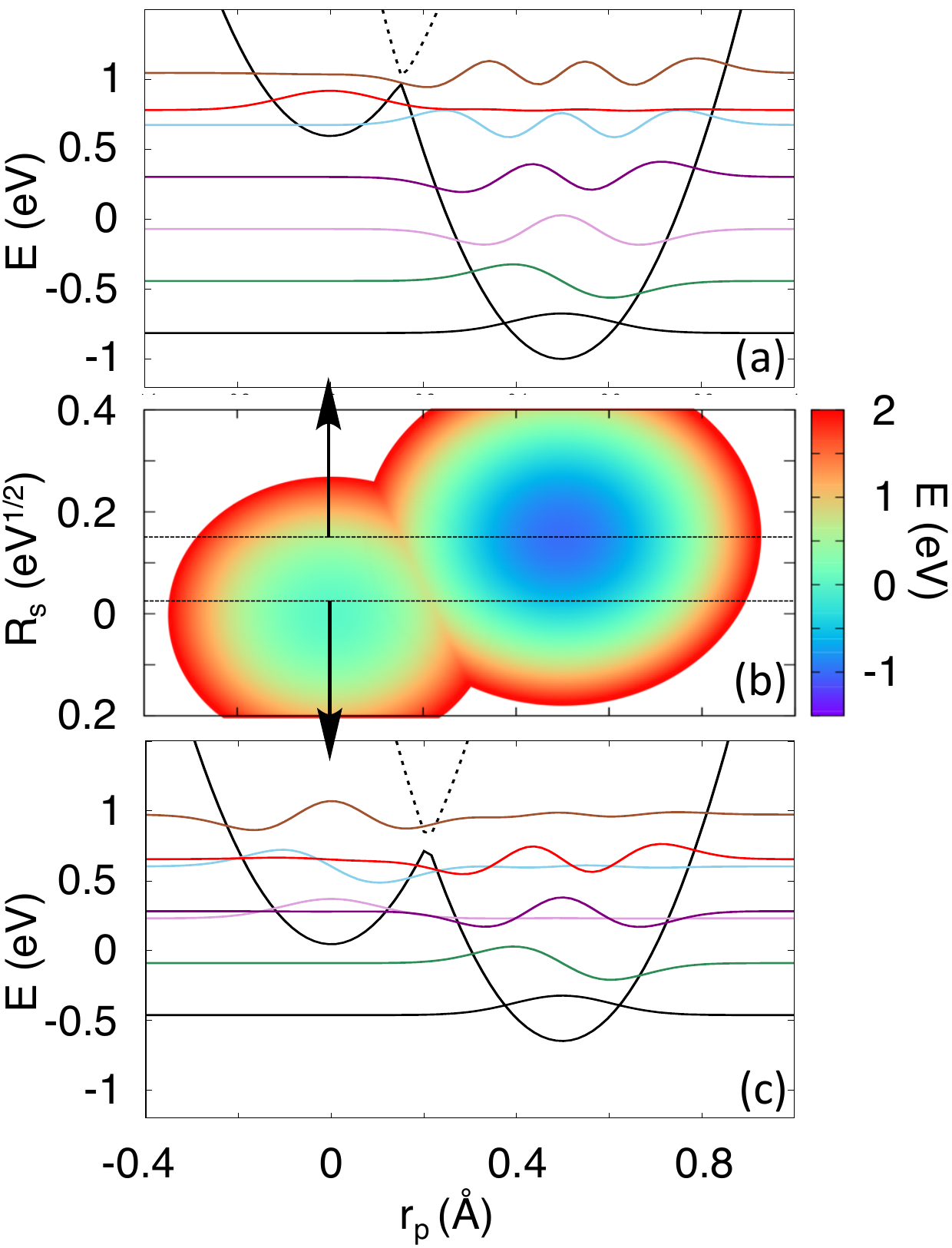}
\caption{The vibronic eigenfunctions with the corresponding eigenvalues for Model IB, as a function of $r_p$ at two different solvent configurations, with (a) right after photoexcitation with $R_s^0=\sqrt{2\lambda/f_0}$, and (c) after solvent relaxes to the donor potential well at $R_s\sim0$. (b) The free energy profile of $\mathrm{S_1}$ state in Model IB as a function of $r_p$ and $R_s$.}
\label{fig:tunneling}
\end{figure}

To demonstrate the effect of proton tunneling on PI-PCET dynamics in Model IB, in Fig.~\ref{fig:tunneling} we present the proton vibrational eigenfunctions $\Psi_{\alpha}(r_\mathrm{p};R_\mathrm{s})=\langle {r_\mathrm{p}}|\Phi_{\alpha}(\hat{r}_\mathrm{p};R_\mathrm{s})\rangle$ with different solvent configurations $R_\mathrm{s}$. In Fig.~\ref{fig:tunneling}a, the solvent configuration $R_s$ is chosen to be at its initial value $R_\mathrm{s}^{0}=\sqrt{2\lambda/f_{0}}$ upon photoexcitation. The initial proton wavepacket is created as a linear combination of vibational states $5-7$, which are blue, red and brown states in Fig.~\ref{fig:tunneling}a. During the first 1 ps, solvent coordinate $R_\mathrm{s}$ has relaxed to the donor side (with $R_{s}\sim0$), with the corresponding vibronic eigenfunctions presented in Fig.~\ref{fig:tunneling}c. At the same time, vibrational relaxation process has induced the population transfer to vibronic states $3-4$ (depicted as the pink and violet states in Fig.~\ref{fig:tunneling}c). Note that there is a sizable barrier along the $\hat{r}_\mathrm{p}$ coordinate for the relaxed solvent configuration ($R_\mathrm{s}\sim0$) in panel (c) compared to panel (a) ($R_\mathrm{s}^{0}=\sqrt{2\lambda/f_{0}}$). Under such circumstances in (c), the populated proton vibrational states (pink and violet) are below the barrier, and the tunneling through the barrier becomes the predominant mechanism for the later stage PI-PCET dynamics. This explains the significant KIE observed in Model IB (Fig.~\ref{fig:KIE}a). In contrast, PI-CPET dynamics in Model IC is dictated by the early stage vibrational relaxation process, with the adiabatic vibronic states that are more similar to Fig.~\ref{fig:tunneling}a, such that it is almost barrier-less for the reaction (akin to the activation less regime in Marcus electron transfer theory), resulting in a negligible KIE. 

\section{Conclusions}
We apply the recently developed QD propagation scheme\cite{Mandal:2018} to investigate the non-adiabatic dynamics of photoinduced proton coupled electron transfer (PI-PCET) reaction. Using PLDM path-integral approach and the electron-proton adiabatic vibronic states as the time-dependent quasi-diabatic states, the outlined QD propagation scheme provides an accurate and efficient theoretical framework for simulating PI-PCET dynamics. Compared to approaches that only treat the transferring electron quantum mechanically but proton classically (such as classical Wigner models\cite{Cotton:2014}), the QD scheme explicitly quantizes proton with vibronic adiabatic states, and thus provides accurate non-adiabatic dynamics and KIE. Compared to the approaches that directly propagate dynamics with the diabatic vibronic basis (such as extended SQC\cite{Kananenka:2018}), the QD scheme only requires a smaller set of adiabatic vibronic states that are directly involved during the non-adiabatic process, thus significantly reduces the computational costs associated with the number of states that need to be explicitly propagated. Previous theoretical studies with diabatic quantum dynamics approaches\cite{Ananth:2017} for simulating PCET reactions usually require parametrizing the original model system into a strict diabatic system-bath model.\cite{Ananth:2017,Kretchmer:2013} This process requires tedious efforts and remains a non-trivial task and a significant challenge for atomistic simulations.\cite{Ananth:2017} Here, by using {\it adiabatic vibronic} basis that can be obtained with routinely available electronic structure calculations,\cite{Goyal:2016} the QD scheme allows directly propagating quantum dynamics with a {\it diabatic} based approach. 

With this QD propagation scheme, we investigate the vibronic population transfer and the KIE of PI-PCET dynamics with various driving forces. For systems with a small driving force, while the vibrational relaxation process significantly impacts the early stage dynamics, solvent relaxation to the donor side of the free energy surface eventually creates a large barrier for PCET, such that proton tunnelling plays a predominant role. A significant KIE will be observed in such scenario. Whereas for a system with a much larger driving force, the vibrational relaxation completely dictates both early stage dynamics as well as longer-time PI-CPET reactions, resulting in a negligible KIE.

Further, we want to emphasize that the outlined QD scheme and the simulation protocol are general enough and not limited to PLDM; they can be directly applied to a wide range of diabatic trajectory-based quantum dynamics approaches. These approaches include but are not limited to symmetrical quasi-classical (SQC),\cite{Cotton:2016} forward-backward quantum-classical Liouville equations (FB-QCLE),\cite{KapralMQCL} generalized quantum master equation (GQME),\cite{Kelly:2015} and quantum-classical path-integral (QCPI) dynamics.\cite{Makri:2015} The QD propagation scheme provides a transformative theoretical framework for studying challenging PI-PCET reactions through accurate diabatic quantum dynamics approaches with efficient adiabatic electronic structure calculations.

Finally, we want to outline three alternative approaches for quantizing proton besides the commonly used vibronic adiabatic states description\cite{Hammes-schiffer:1994} adapted in this study. The first one requires using time-dependent Gaussian basis (TDGB) function to explicitly expand the proton (as well as the other nuclear) wavefunction. This approach that has been utilized in AIMS\cite{Pijeau:2017} or MP/SOFT,\cite{Kim:2009} can be numerically expensive as it requires many TDGB functions associated with each nuclear DOF. The second one uses nuclear-electronic orbital (NEO) approach\cite{Hammes-Schiffer:NEO} in which wavefunctions are used for the transferring electron and proton with molecular orbital techniques, adding additional difficulty on top of the already complex electronic structure problems. The last one quantizes the proton with an imaginary-time path-integral framework\cite{Habershon} in the extended classical phase-space ({{\it i.e.}}, so-called ring polymer). With an explicit description of electronic states, recently emerged state-dependent ring polymer molecular dynamics approaches\cite{Ananth:2013,Richardson:2013,Chowdhury:2017} can potentially provide accurate electronic non-adiabatic dynamics with nuclear quantum effects, and thus are promising for investing PI-PCET dynamics once combined with the QD-propagation scheme.

\section {Acknowledgement}
{This work was supported by the University of Rochester startup funds. Computing resources were provided by the Center for Integrated Research Computing (CIRC) at the University of Rochester. FAS appreciates valuable discussions with Prof. Anirban Hazra.}

\section{Appendix A: Solvent parameters for Model I}
We provide the details of the parameters used in Model I. The force constant for the collective solvent DOF (so-called ``inverse Pekar factor") is $f_0=4\pi \epsilon_0 \epsilon_{\infty}/(\epsilon_0 - \epsilon_{\infty})$, where $\epsilon_0$ and $\epsilon_{\infty}$ are the inertial and optical dielectric constants characterizing the polarizability of the solvent. Here, we chose these parameters that correspond to water as the solvent.\cite{Hazra:2011} 

\begin{table}[htbp]\label{parameter}
\caption{Parameters used in Langevine dynamics.}
\begin{tabular*}{0.35\textwidth}{@{\extracolsep{\fill} } l c }
\hline
{\small Parameter} & {\small Water at 298 K} \\
\hline
$\epsilon_0$ & $79.2$ \\
$\epsilon_{\infty}$ & $4.2$ \\
$f_0$ & $55.7$ \\
$\tau_0$ (ps) & $0.0103$ \\
$\tau_{\text{D}}$ (ps) & $8.72$ \\
$M_s$ ($\text{ps}^2$) & $0.265$ \\
$\lambda$ (eV) & $0.65$ \\
\hline
\end{tabular*}
\end{table}

Further, $\tau_{\text{L}}=\epsilon_{\infty}(\tau_0+\tau_{\text{D}})/\epsilon_0$ is the longitudinal relaxation time accounting for the long-time solvent response function, where $\tau_{\text{D}}$ is the Debye relaxation time and $\tau_0$ is the characteristic rotational time of the solvent molecules. All of the parameters used in this paper are tabulated in Table I and a full description of them could be found in Ref.~\citenum{Hazra:2011}.

\section{Appendix B: Hamiltonian for Model II}
Here we provide the details for the PI-PCET model that does not contain collective solvent coordinate,\cite{Venkataraman:2009,Cotton:2014} which is referred to as Model II in this paper. This Hamiltonian is used to explore the accuracy of various recently developed non-adiabatic approaches as well as the role of proton quantization for PI-PCET reactions in this study. Note that only the results presented in Fig.~\ref{fig:vib} are obtained with this model system. 

The total Hamiltonian is defined in Eqn.~\ref{eqn:totham}. The electron-proton Hamiltonian $\hat{H}_\mathrm{ep}$ is expressed as following
\begin{eqnarray}\label{modelH}
\hat{H}_\mathrm{ep}=\hat{T}_\mathrm{p}+
\Bigg [
  \begin{tabular}{cc}
  $U^\mathrm{D}(r_\mathrm{p})$ & $V_\mathrm{DA}$ \\
   $V_\mathrm{DA}$ & $U^\mathrm{A}(r_\mathrm{p})$\\
  \end{tabular}
\Bigg ]
\end{eqnarray} 
Here $r_\mathrm{p}$ is the proton coordinate, $U^\mathrm{D}(r_\mathrm{p})$ and $U^\mathrm{A}(r_\mathrm{p})$ are the proton potential associated with electronic donor and acceptor states that have exactly the same expression as Model I (see Eqn.~\ref{eqn:donor-accep}), with $r_\mathrm{p}^\mathrm{D}=0$ and $r_\mathrm{p}^\mathrm{A}=-0.5$ \AA. Prior to photoexcitation, proton is on the vibrational ground state of the $\mathrm{S_0}$ state, with the potential $U^0=\frac{1}{2}m_\mathrm{p}\omega_\mathrm{p}^{2} (r_\mathrm{p}-r^{0}_\mathrm{p})^2$ where $r^{0}_\mathrm{p}=-0.15$ \AA. The rest of parameters in the above Hamiltonian have the same values as used in Model IA, with $\Delta=0$ eV, $V_\mathrm{DA}=0.03$ eV, $\omega_{p}=3000~\mathrm{cm}^{-1}$, and $m_\mathrm{p}=1.0073$ amu.

The bath Hamiltonian which describes the interaction between the electron-proton system and a condensed-phase solvent environment is modeled by coupling of the donor electronic state to a dissipative harmonic bath with the following expression
\begin{equation}
H_\mathrm{sb}= \sum_{k=1}^{K}\left[ \frac{P_{k}^2}{2M_{k}} + \frac{1}{2}M_{k} \omega_{k}^{2} \left(R_{k}- \frac{c_{k}}{M_{k} \omega_{k}^{2}}|D\rangle\langle D|\right)^2\right],
\end{equation}
where $R_{k}$ and $P_{k}$ represent the $k^\mathrm{th}$ bath coordinate and momentum, with $M_k$ and $\omega_k$ as the corresponding mass and frequency. The bath is characterized by an Ohmic spectral density $J(\omega) =  \frac{1}{2}\pi\xi\omega e^{-\omega/\omega_\mathrm{c}}$, where  $\xi$ is the unit-less Kondo parameter and $\omega_c$ is the cut-off frequency. Here, we use $\xi=24$ and $\omega_\mathrm{c}=600~\mathrm{cm^{-1}}$. Discretizing this spectral density yields $N$ harmonic oscillators with frequencies $\omega_k = -\omega_c \ln\big(1-k\frac{\omega_0}{\omega_c}\big)$ and coupling constants, $c_k= \sqrt{\xi\omega_0M_k}\omega_k$. Here, $\omega_0$ for a total of $K$ bath modes is given by $\omega_0 = \frac{\omega_c}{K}\big(1- e^{-\omega_\mathrm{m}/\omega_c} \big)$,  and $\omega_\mathrm{m}$ was chosen to be $3\omega_c$.

The initial condition for the PLDM simulation are provided as follows. The initial conditions for the bath modes are sampled from the Wigner distribution for harmonic oscillators' thermal density as follows
\begin{equation}
\rho_\mathrm{b}^\mathrm{W} = \Pi_{k=1}^{K} { \omega_k \Gamma_k} e^{-\Gamma_{k} \big[{{P_{k}^2\over {2M_k}} + {1\over2}M_k\omega_{k}^2({R_k}-{R_k^0})^2\big]}},
\end{equation}
where $\Gamma_{k} = (2/\omega_k)\tanh(\omega_k/2k_\mathrm{B}T)$, and $\omega_k$ sampled from the spectral density, and $R_k^0=c_{k}/(M_{k} \omega_{k}^{2})$. Further, we use the focused initial conditions\cite{FocusedBonella} to facilitate the convergence of the sampling for the mapping variable.  

For the el-PLDM calculation, we choose to treat proton classically, with the corresponding initial conditions sampled from the following function
\begin{equation}
\rho_\mathrm{p}^\mathrm{W} = { \omega_\mathrm{p} \Gamma_\mathrm{p}} e^{-\Gamma_\mathrm{p} \big[T_\mathrm{p} + {1\over2}m_\mathrm{p}\omega_\mathrm{p}^2({r_\mathrm{p}}-{r_\mathrm{p}^0})^2\big]},
\end{equation}
where $\Gamma_\mathrm{p} = (2/\omega_p)\tanh(\omega_p/2k_\mathrm{B}T)$, $T_\mathrm{p}$ is the classical kinetic energy of the proton, $r_\mathrm{p}$ is the proton coordinate, $\omega_\mathrm{p}$ is the proton vibrational frequency.

The converged results for Model II with el-PLDM method are obtained by propagating an ensemble of $10^4$ trajectories, with a time step of $dt=0.024$ fs. For the same model, we use 2000 trajectories for vib-PLDM or QD-PLDM propagation. The total number of vibrational basis $\{|\phi_\mathrm{p}^{m}\rangle\}$ used in this model is 80. In the QD-PLDM propagation, we use the first 20 low-lying time-dependent vibronic states as the QD basis. 

\bibliography{pipcet}
\end{document}